\documentclass[12pt,preprint]{aastex}

\newcommand{\asca}{{\it ASCA}}
\newcommand{\sax}{{\it BeppoSAX}}
\newcommand{\xmm}{{\it XMM-Newton}}

\shorttitle {The Effects of Periodic Data Gaps on CCF Lag Determinations}
\shortauthors {Zhang et al.}

\begin{document}

\title {THE EFFECTS OF PERIODICALLY GAPPED TIME SERIES ON CROSS-CORRELATION 
	LAG DETERMINATIONS}

\author {Y.H. Zhang}
\affil  {Center for Astrophysics, Department of Physics, Tsinghua University, 
100084 Beijing, P.R. China, and Dipartimento di Scienze, Universit\`a degli 
Studi dell'Insubria, via Valleggio 11, I-22100 Como, Italy} 
\email  {youhong.zhang@mail.tsinghua.edu.cn; youhong.zhang@uninsubria.it}
\author {I. Cagnoni, A. Treves} 
\affil	{Dipartimento di Scienze, Universit\`a degli Studi dell'Insubria, 
	 via Valleggio 11, I-22100 Como, Italy} 
\author {A. Celotti}
\affil  {International School for Advanced Studies, SISSA/ISAS,
   	 via Beirut 2-4, I-34014 Trieste, Italy}
\author {L. Maraschi}
\affil  {Osservatorio Astronomico di Brera, via Brera 28,
   	 I-20121 Milano, Italy}

\begin{abstract}

The three bright TeV blazars Mrk~421, Mrk~501 and PKS~2155--404 are
highly variable in synchrotron X-ray emission. In particular, these sources
may exhibit variable time lags between flux variations at different X-ray
energy bands. However, there are a number of issues that may significantly
bias lag determinations. Edelson et al. (2001) recently proposed that the
lags on timescales of hours, discovered by \asca\ and \sax , could be an
artifact of periodic gaps in the light curves introduced by the Earth
occultation every $\sim$ 1.6 hr. Using Monte Carlo simulations, in this
paper we show that the lags over timescales of hours can not be the
spurious result of periodic gaps, while periodic gaps indeed introduces
uncertainty larger than what present in the evenly sampled data. The results
also show that time lag estimates can be substantially improved by using
evenly sampled light curves with large lag to bin-size ratio. Furthermore,
we consider an \xmm\ observation without interruptions
and re-sample the light curves using the \sax\ observing windows, and then
repeat the same cross correlation function (CCF) analysis on both the real
and fake data. The results also show that periodic gaps in the light curves
do not significantly distort the CCF characters, and indeed the CCF peak ranges
of the real and fake data overlap. Therefore, the lags
discovered by \asca\ and \sax\ are not due to periodic gaps in the
light curves.

\end{abstract}



\section{Introduction}\label{sec:intro}

One of the main advances from recent X-ray observations of blazars is the
discovery of the energy-dependent time lags in the X-ray emission of the
three bright TeV-emitting blazars Mrk~421, Mrk~501 and PKS~2155--304.
The overall spectral energy distributions (SEDs) show that the synchrotron
emission component from these sources peaks at high-energy (UV/soft X-ray)
band. This indicates that the X-ray emission from these sources is the high
energy tail of the synchrotron component, where the most violent
variability is expected. The inter-band time lag is one of the important
variability parameters. The cross-correlation function (CCF)
technique is the standard tool for lag determinations. \asca\ and \sax\
discovered that in these sources the lower energy X-ray photons may lag or
lead the higher energy ones, i.e., the so-called soft or hard lag,
respectively. The signs and values of the lags may well depend on either
energy or on the single flare analyzed. The typical lags (either soft or
hard lag) range from $\sim$ zero to $10^{4}$~s (see Zhang et al. 2002 for a
review), and appear to be correlated to the flare duration: the
shorter the flare duration, the smaller the lag (Zhang et al 2002;
Brinkmann et al.  2003). Zhang (2002) also found evidence for a
dependence of lags on timescales in Mrk~421: lags appear to be larger on
longer timescales.  The observed lags have been interpreted as
evidence for the interplay of the acceleration and cooling timescales of
relativistic electrons responsible for the observed X-rays taking place in
the jets (e.g., Kirk, Rieger \& Mastichiadis 1998), and used to constrain
the main parameters of the emitting region on the basis of variability 
(e.g., Zhang et al. 2002) rather than of SEDs, as commonly used.

However, the latest lag searches with \xmm\ have put the discovery of \asca\ 
and \sax\ into question. Edelson et al. (2001) and Sembay et
al. (2002) reported that there is no evidence for measurable inter-band
lags in the X-rays, with upper limits of $\sim$ 0.3 (PKS~2155--304) and
0.08 hr (Mrk~421). It is commonly believed that uninterrupted time series,
due to the highly eccentric $\sim$~48 hours orbit of \xmm , allow to
detect more reliably the lags. This has led these authors to propose that
the lags measured with \asca\ and \sax\ could well be an artifact of the
periodic interruptions by the Earth occultation related to the short \asca\
and \sax\ orbital period ($\sim 1.6$ hr). Maraschi et al. (2002) also
reported no measurable lags with a full orbit \xmm\ observation of
PKS~2155--304. Brinkmann et al. (2003) re-analyzed all the available
\xmm\ observations of Mrk~421, working with light curves that have a small
bin-size of 8~s, and found typical lags of $\sim5$~min.

In order to address the issue of the reliability and significance of the
lags discovered by \asca\ and \sax , in this paper we perform Monte Carlo
simulations to investigate the effects of periodically gapped time series
on CCF lag determinations. A number of simulations using inverse Fourier
transformation, which is model-dependent, have been carried out in the
literature to evaluate the significance and uncertainty of the CCF lag
determinations either for poorly sampled time series (e.g., Gaskell \&
Peterson 1987; Maoz \& Netzer 1989; White \& Peterson 1994) or for evenly
sampled time series (e.g., Welsh 1999). The issue of poorly irregular
sampling and noisy data was also previously investigated (see the summary
in Koen 1994). There are a number of issues that can significantly bias the
reliability of CCF lag determinations.  We refer to Welsh (1999) for a
review on such issue, where the author also showed the effects of power
spectral density (PSD) slope, duration, signal-to-noise ratio, de-trending
and tapering of light curves.  We note, however, that the previous
simulations mainly dealt with the relationship between the UV/optical
emission line flux and the continuum variations in Seyfert galaxies, which
applied to the reverberation mapping (to measure the size of
broad-line region) and typical ground-based irregular sampling patterns. In
the blazar context, the simulations by Litchfield, Robson, \& Hughes (1995)
were performed on the radio sampling patterns. Our simulations are
tailored to match the sampling characteristics of space-based observations
introduced by low earth orbit satellites, which to our knowledge have not
been explicitly addressed in astronomical literature. 
Another distinction is that we discuss the time lags between the
variations of the synchrotron emission in different X-ray bands. In
particular, it is worth noting that there are obvious difficulties in
accurately determining lags in the synchrotron X-ray emission of TeV
blazars as (1) the lags are short compared to the bin-sizes of the
available data; (2) the data are usually not equally sampled -- in
particular periodical interruptions of space-based observations are 
unavoidable for the low Earth orbit satellites; and (3) there are 
ambiguities in interpreting the complexities of the CCF results. 

In this work we also make use of the method introduced by Peterson et
al. (1998), namely a model-independent Monte Carlo method to assess the
uncertainties in the lag measurements obtained. This method does make use
of real data, and is known as flux redistribution/random subset selection
(FR/RSS) that is based on the ``bootstrap'' method. Here we adopt it 
to study the effects of periodic gaps in lag determinations. To do so,
we use a real \xmm\ observation without interruptions, and re-sample
it with the typical \sax\ sampling windows. The same CCF analysis is 
then performed on both the real and fake time series.

In \S\ref{sec:simu} we conduct model-dependent simulations: 
\S\ref{sec:simu:assum} illustrates the assumptions for the simulations to
be performed; and the results are presented in
\S\ref{sec:simu:results}. The model-independent simulations are performed
in \S\ref{sec:xmm}. We discuss the significance of our results in
\S\ref{sec:disc}.


\section{Monte Carlo Simulations} \label{sec:simu}

\subsection{Assumptions}\label{sec:simu:assum}

The inverse Fourier transformation from frequency to time domain is
usually performed to simulate light curves by assuming a PSD model. We use
the algorithm of Timmer \& K\"{o}nig (1995), which randomizes both the
amplitude and the phase of a red noise process at each Fourier
frequency. In exploring the full variety of possible light curves showing
the same PSD, this algorithm is superior to the commonly used one that
randomizes phases only (Benlloch et al. 2001; Uttley, M${\rm ^c}$Hardy, \&  
Papadakis 2002). In order to mimic a real situation as
closely as possible, we normalize fake light curves on the basis of 
the real variability behavior of one observation of Mrk~421 obtained
with \sax\ (Fossati et al. 2000; Zhang 2002) as this observation has 
a good signal-to-noise ratio that will lay stress on the key point of this
investigation. The assumptions and procedure of our simulations are as
follows.

A simple power-law PSD, i.e., $P(f) \propto f^{-\alpha}$ with slope
$\alpha=2.5$ is assumed (Zhang 2002) to represent the red noise variability
of the bright TeV blazars (Kataoka et al. 2001; Zhang et al. 1999; 2002), 
and from this the light curves are recovered on the basis of  
Monte Carlo technique. With one set of Gaussian distributed random 
numbers, we construct a fake pair of light curves that evenly sample 171
points with bin-size $\Delta t = 512$~s.  This choice resembles the binned
light curves of Mrk~421 used in Zhang (2002) to perform the CCF
analysis. At the same time, in order to mimic the discovered time lags, we
delay the phase of the second light curve of the fake pair on the basis of
the relationship $\Delta \phi (f_i) = 2\pi f_i\tau(f_i)$, where $f_i$ is
the Fourier frequency ($i=1,..., N/2$; $N=171$ is the total number of the
evenly sampled light curve points), $\tau(f_i)$ is the time lag and $\Delta
\phi (f_i)$ the phase lag at $f_i$.  For simplicity, we assume that $\tau$
is frequency-independent, i.e., $\tau=$ constant (see, however, Zhang 2002
for the evidence of the dependence of $\tau$ on frequency), and in turn
$\Delta \phi (f_i) = 2\pi\tau f_i$.  An assumed (true) lag $\tau$ therefore
corresponds to the lag determined with the CCF methods. The aim of our
simulations is to recover the assumed $\tau$ by applying the CCF methods to
the fake pair with specific sampling windows. The two light curves of the
fake pair are then scaled to have the same mean and variance as the real
0.1--2~keV and 2--10~keV light curves of Mrk~421 (Zhang 2002),
respectively. In order to mimic photon counting (Poisson) white noise, the
two fake light curves are further Gaussian randomly redistributed on the
basis of the average errors on the real 0.1--2~keV and 2--10~keV light
curves of Mrk~421. In order to simulate the periodic gaps of the light 
curves obtained with \sax , we re-sample 
the two fake light curves by applying the real observing windows on the
0.1--2~keV and 2--10~keV light curves of Mrk~421. Three CCF methods,
i.e., interpolation cross correlation function (ICCF, White \& 
Peterson 1994), Discrete Correlation Function (DCF, Edelson \& Krolik
1998), and \textit{Fisher's z-transformed} DCF (ZDCF, Alexander 1997) that
is based on the DCF, are then used to cross-correlate the two fake light
curves before and after applying the real observing windows. All CCFs are
normalized by the mean and standard deviation of the two cross-correlated
light curves using only the data points that actually contribute to the
calculation of each lag (White \& Peterson 1994), since the light curves in
our cases are not stationary.  A simulation is deemed to have succeeded if
$r_{\rm max}$ (the maximum value of the CCF) between the two fake light
curves is significant at a level of confidence greater than $95\%$
(Peterson et al. 1998). For each succeeded trial, we record the lags using
three techniques to interpret the CCF results: (1) using the lag
corresponding to $r_{\rm max}$ of the CCF, $\tau_{\rm peak}$; (2) computing
the centroid of the CCF over time lags bracketing $r_{\rm max}$, $\tau_{\rm
cent}$  -- all the CCF points with $r$ in excess of 0.8$r_{\rm max}$
are used (Peterson et al. 1998); (3) fitting the CCF with a Gaussian
function to find the location of the CCF peak, $\tau_{\rm fit}$. We repeat
this procedure 2000 times to construct probability distributions of the lags 
(i.e., the cross-correlation peak distribution, CCPD; Maoz \& Netzer 1989)
for $\tau_{\rm peak}$, $\tau_{\rm cent}$, and $\tau_{\rm fit}$.

\subsection{Results}\label{sec:simu:results}

We simulate time lags covering the range discovered by ASCA and \sax\ 
in TeV blazars. Therefore, we arbitrarily assume $\tau = 0, 300, 1400,
3000, 5400, 7100$~s, respectively. We create fake light curves using the
same sets of random number for each case except for different $\tau$.

\begin{figure}
\epsscale{1.0}
\plotone{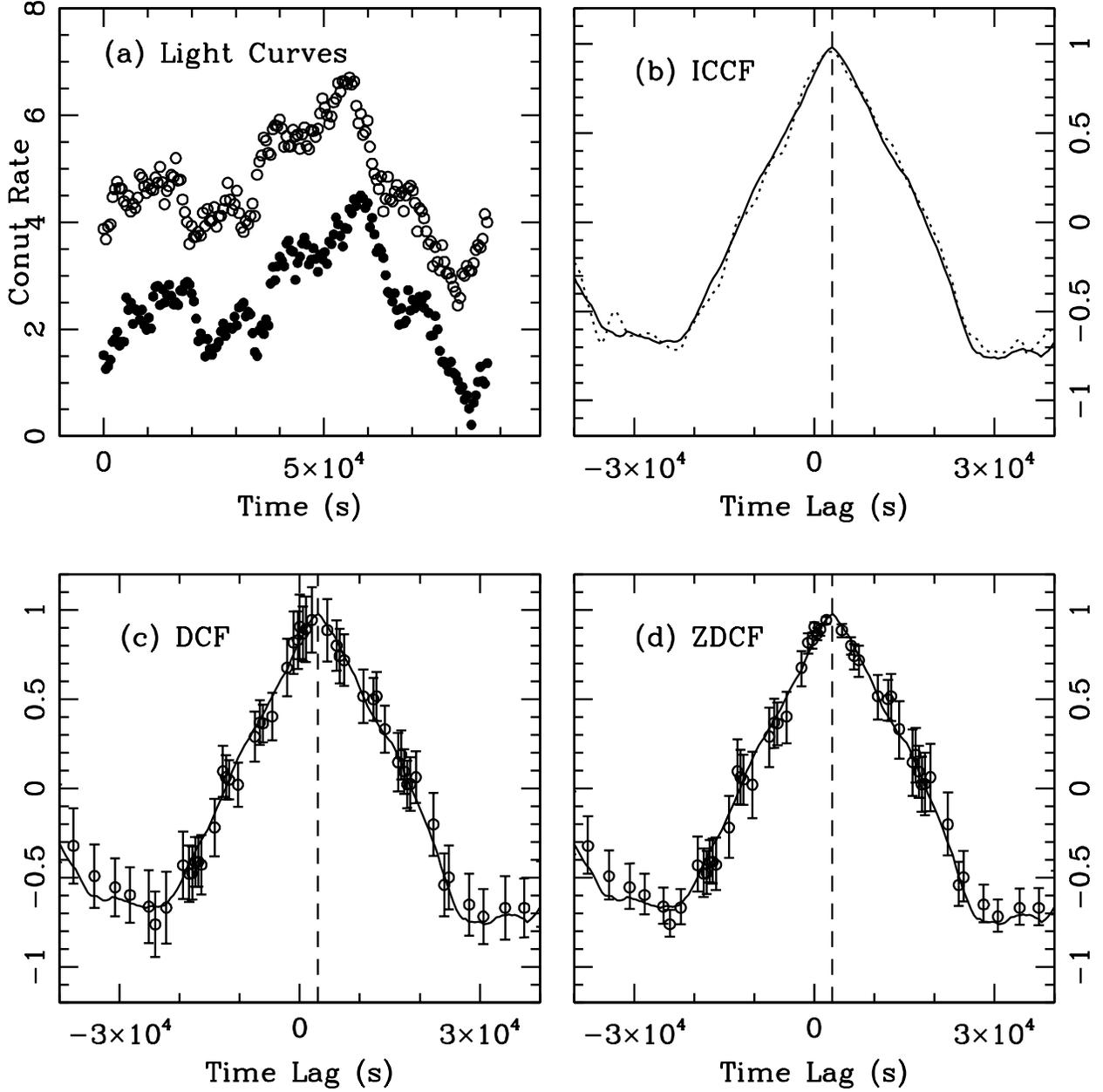}
\caption {\footnotesize (a) An example of a fake light curve pair
assuming $\tau=3000$~s between them. For clarity, the count rates of the
second light curve (solid circles) are lowered by 1.5.  (b) the
corresponding ICCF of (a). The solid line is obtained from evenly sampled
light curves, and the dashed line is obtained from periodically gapped
light curves re-sampled from (a) after applying the real \sax\ observing
windows; (c) the corresponding DCF of (a). The solid line is obtained from
evenly sampled light curves, and the open circles with error bars are
obtained from the same periodically gapped light curves as in (b);
(d) same as (c), but for the ZDCF.  The dashed lines in (b), (c) and
(d) indicate the true lag.}
\label{fig:lc}
\end{figure}

Figure~\ref{fig:lc}a, as an example, shows a pair of fake light curves that
are evenly sampled, and the second light curve (solid circles) is delayed
by $\tau=3000$~s with respect to the first one (open circles). The lag is
clearly visible by comparing both the peaks and the troughs of the two
light curves. In this case, the ratio of $\tau$ to $\Delta t$ of the light
curves is $\sim 6$. The ICCF, DCF and ZDCF of the two light curves are
shown in Figure~\ref{fig:lc}b-d (solid lines). We use $\tau_{\rm peak}$,
$\tau_{\rm cent}$, and $\tau_{\rm fit}$ to measure the lag,
respectively. The results are reported in Table~\ref{tab:simccf}. One
can see that the true lag is properly recovered by the 
different CCF methods and different techniques used to interpret the CCFs.

We then re-sample these two light curves using the real \sax\ observing
windows as mentioned in \S\ref{sec:simu:assum}. The resulted light curves
thus are affected by periodic gaps resembling the light curves 
obtained with \sax .  Note that the number of points (63) in the first
re-sampled light curve is smaller than that in the second one (76) because
on-board \sax\ the LECS detector is less exposed than the MECS one. The
ICCF, DCF and ZDCF of the two light curves with periodic interruptions are
shown in Figure~\ref{fig:lc}b-d (dotted line or open circles with error
bars).  The measured lags are also reported in
Table~\ref{tab:simccf}. In this case, both $\tau_{\rm peak}$ and $\tau_{\rm
cent}$ underestimate the true lag while $\tau_{\rm fit}$ of the DCF
and ZDCF still recover it, although with larger uncertainties. 
Figure~\ref{fig:lc}b-d also show that the overall
characteristics of the CCFs obtained from the evenly sampled light
curves are almost identical to those obtained from the corresponding
light curves with periodic gaps. However, due to the missing of a
large number of data points in the latter case, the CCF peaks somewhat
shift to smaller lags. Note that $\tau_{\rm fit}$ is not calculated
for the ICCF case because there are no estimates of the ICCF errors, 
and the DCF errors are overestimated with respect to the ZDCF ones.

\begin{figure}
\epsscale{1.0}
\plotone{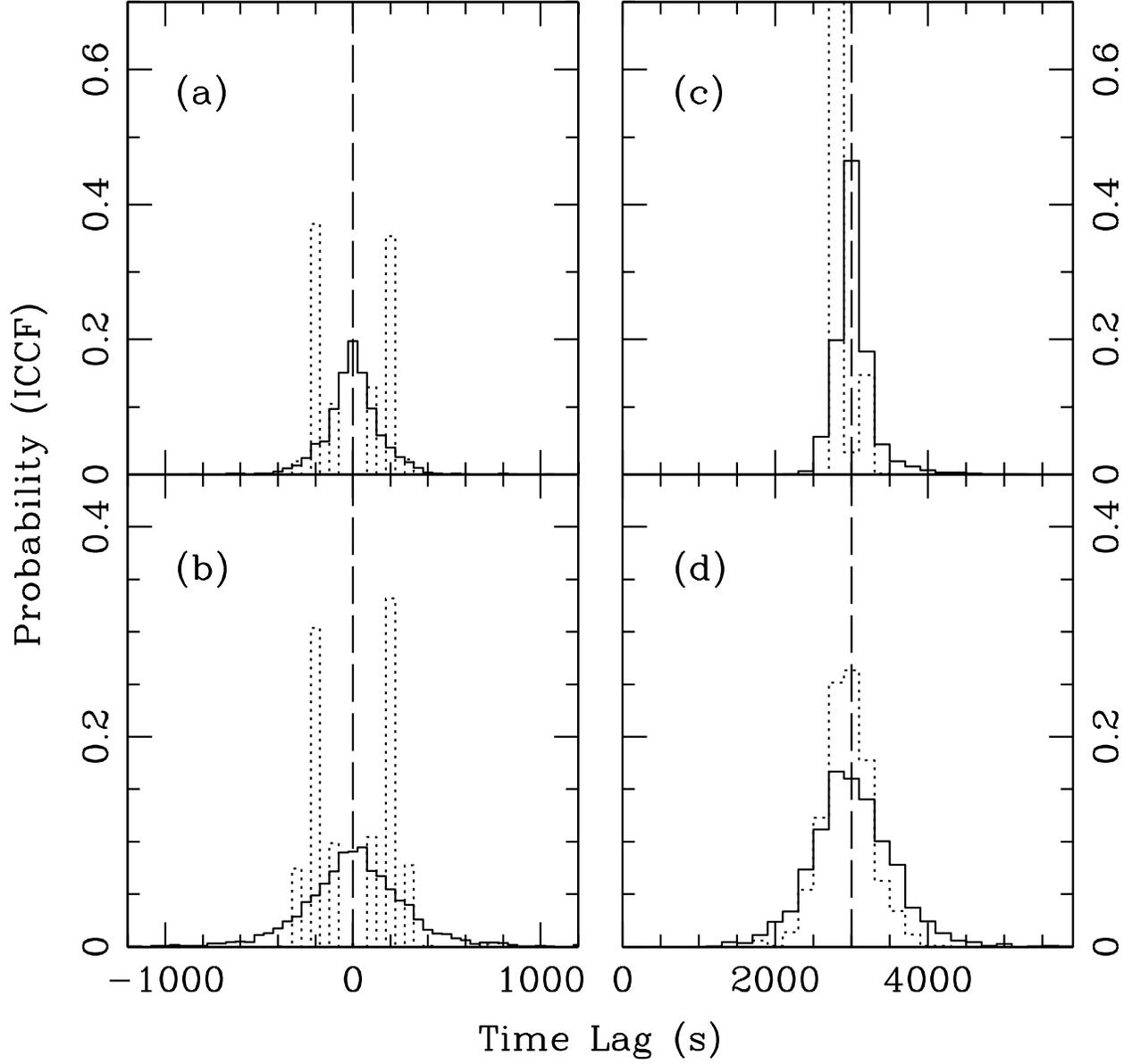}
\caption{ \footnotesize  
ICCF CCPDs for two different lags. 
(a) evenly sampled light curves with $\tau=0$~s; 
(b) periodically gapped light curves with $\tau=0$~s; 
(c) evenly sampled light curves with $\tau=3000$~s; 
(d) periodically gapped light curves with $\tau=3000$~s.
The dotted line refers to $\tau_{\rm peak}$, and the solid line to 
$\tau_{\rm cent}$. 
The vertical long dashed line indicates the true lag.
}
\label{fig:prob1}
\end{figure}

\begin{figure}
\epsscale{1.0}
\plotone{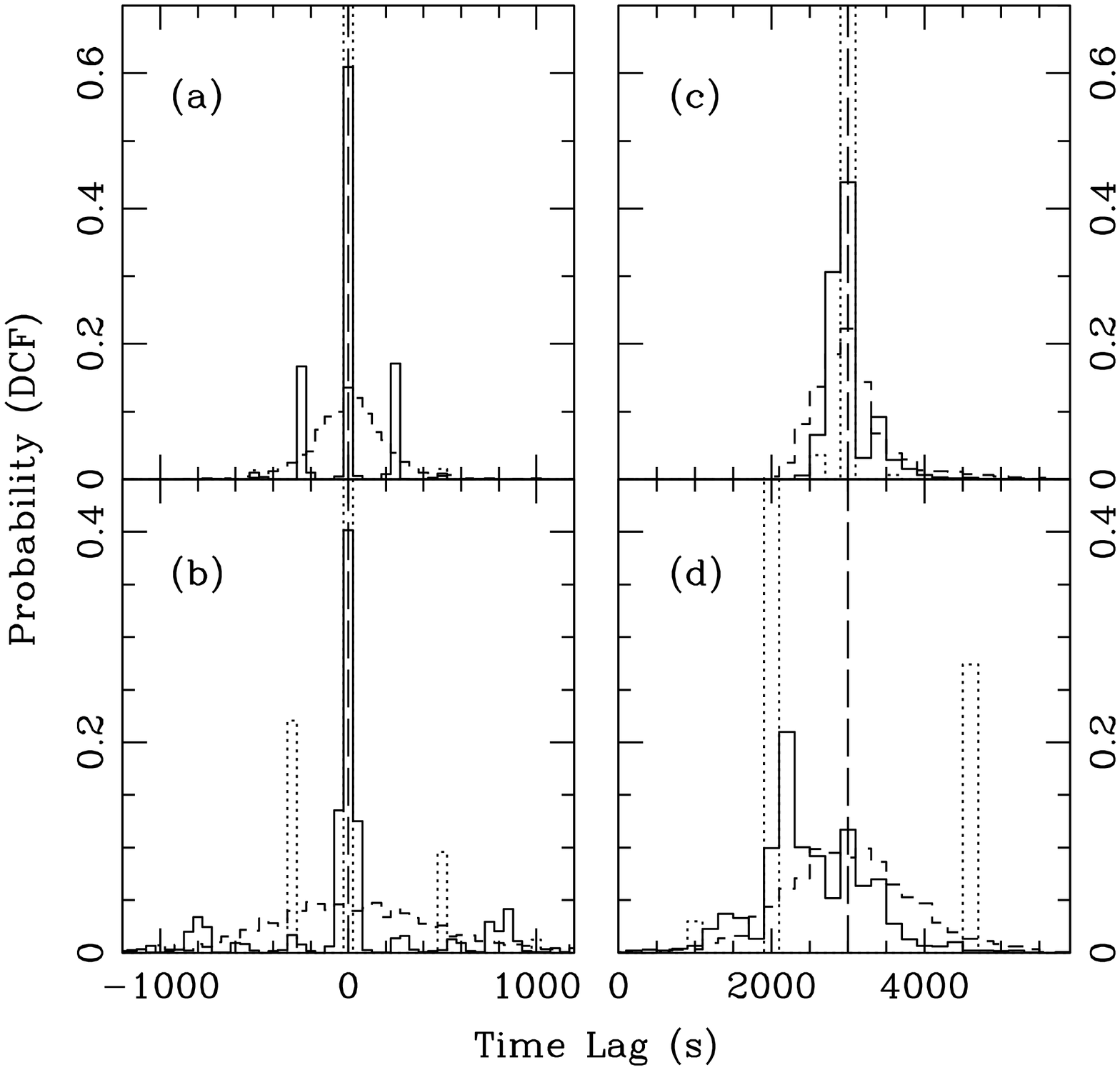}
\caption{ \footnotesize  
Same as Figure~\ref{fig:prob1}, but for DCF.
The dotted line refers to $\tau_{\rm peak}$, the solid line to 
$\tau_{\rm cent}$, and the short dashed line to $\tau_{\rm fit}$. 
The vertical long dashed line indicates the true lag.
}
\label{fig:prob2}
\end{figure}

\begin{figure}
\epsscale{1.0}
\plotone{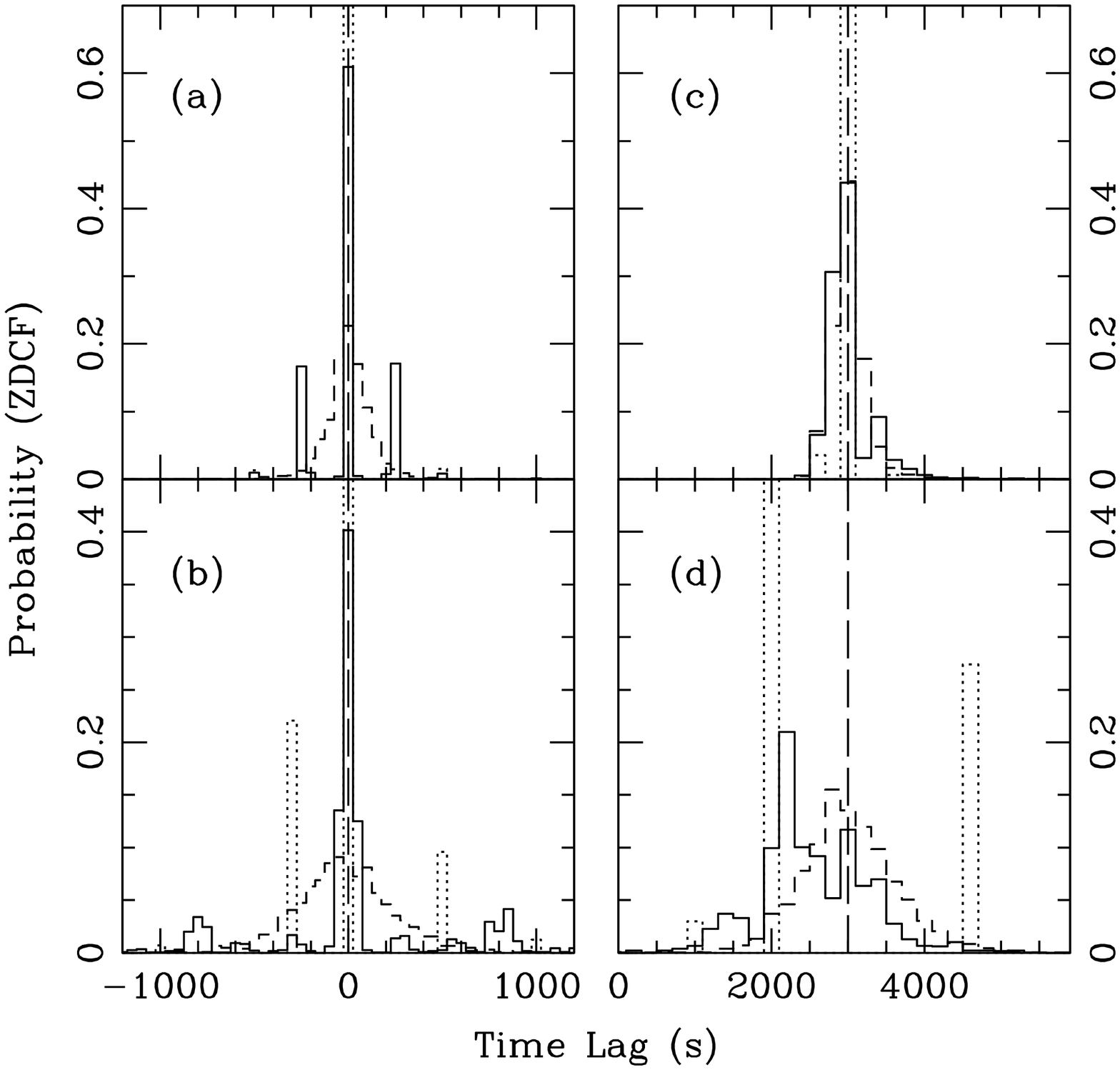}
\caption{ \footnotesize  
Same as Figure~ \ref{fig:prob1}, but for ZDCF.
The dotted line refers to $\tau_{\rm peak}$, the solid line to 
$\tau_{\rm cent}$, and the short dashed line to $\tau_{\rm fit}$. 
The vertical long dashed line indicates the true lag.
}
\label{fig:prob3}
\end{figure}

The statistical significance of the effects of periodic gaps on the 
CCF lag determinations is deduced from the probability distributions of a
number of simulations. We show in Figure~\ref{fig:prob1} (ICCF),
Figure~\ref{fig:prob2} (DCF), and Figure~\ref{fig:prob3} (ZDCF) the CCPDs
of the simulated lags for the case $\tau = 0$ and $3000$~s. It 
is worth noting that the CCPDs are almost always non-normal distributions,
in particular for $\tau_{\rm peak}$ and $\tau_{\rm cent}$. Therefore, we use 
the median and $68\%$ confidence
level (with respect to the median) to statistically characterize the
CCPDs. The statistical results for all the simulations performed are
tabulated in Table~\ref{tab:simccpd}. The first column is the assumed true
lag, and columns 2--4 and 5--7 give the CCPD median and $68\%$ confidence
range for evenly sampled and periodically gapped light curves,
respectively. As shown in Figures~\ref{fig:prob1}-\ref{fig:prob3} and
Table~\ref{tab:simccpd}, the main results of our simulations can be
summarized as follows: (1) the assumed true lags are recovered in all cases
in terms of the CCPD medians with $68\%$ confidence errors; (2) the main 
effect of periodic gaps is to broaden the CCPDs, thus to increase the 
uncertainty of the CCF lag determinations. Most importantly, periodic gaps 
do not produce artificial CCF lags. More specifically, from the
simulations with true small lags (e.g, the cases of 0 and 300~s lags), one
can see that periodic interruptions in the light curves definitely do not
produce spurious lags on timescales of hours; (3) in some cases, the three
CCF methods and the three techniques used to interpret the CCF results do
not give rise to completely consistent results. For example, $\tau_{\rm peak}$
strongly depends on the lag steps used to calculate the CCF; the errors of
the DCF are overestimated, thus producing the broadest CCPDs.

These results imply that (1) evenly sampled light curves with large
lag to bin-size ratios give more reliable CCF lag determinations; (2) more
importantly, the lags on timescales of hours discovered by \asca\ and \sax\
can not be an artifact of periodic gaps in the light curves that have
intrinsically small lag; (3) the only effect of periodic gaps is to
introduce uncertainty and to increase the variance on the CCF lag
determinations.


\section{A Specific Case: \xmm\ Observations of PKS~2155--304} \label{sec:xmm}

The simulations presented in the previous section showed that
periodical gaps in the light curves do not produce spurious lags between them, 
but only increase the uncertainty and variance in lag
determinations. Uninterrupted data with high temporal resolution are
available from \xmm\ observations of Mrk~421 and PKS~2155--304. All of
the CCF analysis, performed by Brinkmann et al. (2001; 2003),
Edelson et al. (2001), Sembay et al. (2002), and Maraschi et al. (2002)
showed that the inter-band lags between the soft and hard energy band
are close to zero, with upper limits of about 1000~s. As we
pointed out in the Introduction, these results led Edelson et
al. (2001) to suggest that previous claims of time lags on time scales of
hours might be an artifact of the periodic interruptions every
$\sim$1.6 hours due to the low-Earth orbits of satellites such as
\asca\ and \sax . Therefore, in addition to the simulations presented
in the previous section arguing against the above suggestion, an important
test to assess the role of periodic gaps is to consider the \xmm\ data 
and re-sample them according to the \asca\ or \sax\ observing windows, 
and then repeat the CCF analysis on the fake \asca\ or \sax\
data. Whether or not inter-band lags would be detected from the fake
data would be a strong argument in favor or against the claim by Edelson 
et al. (2001).

\begin{figure}
\epsscale{1.0}
\plotone{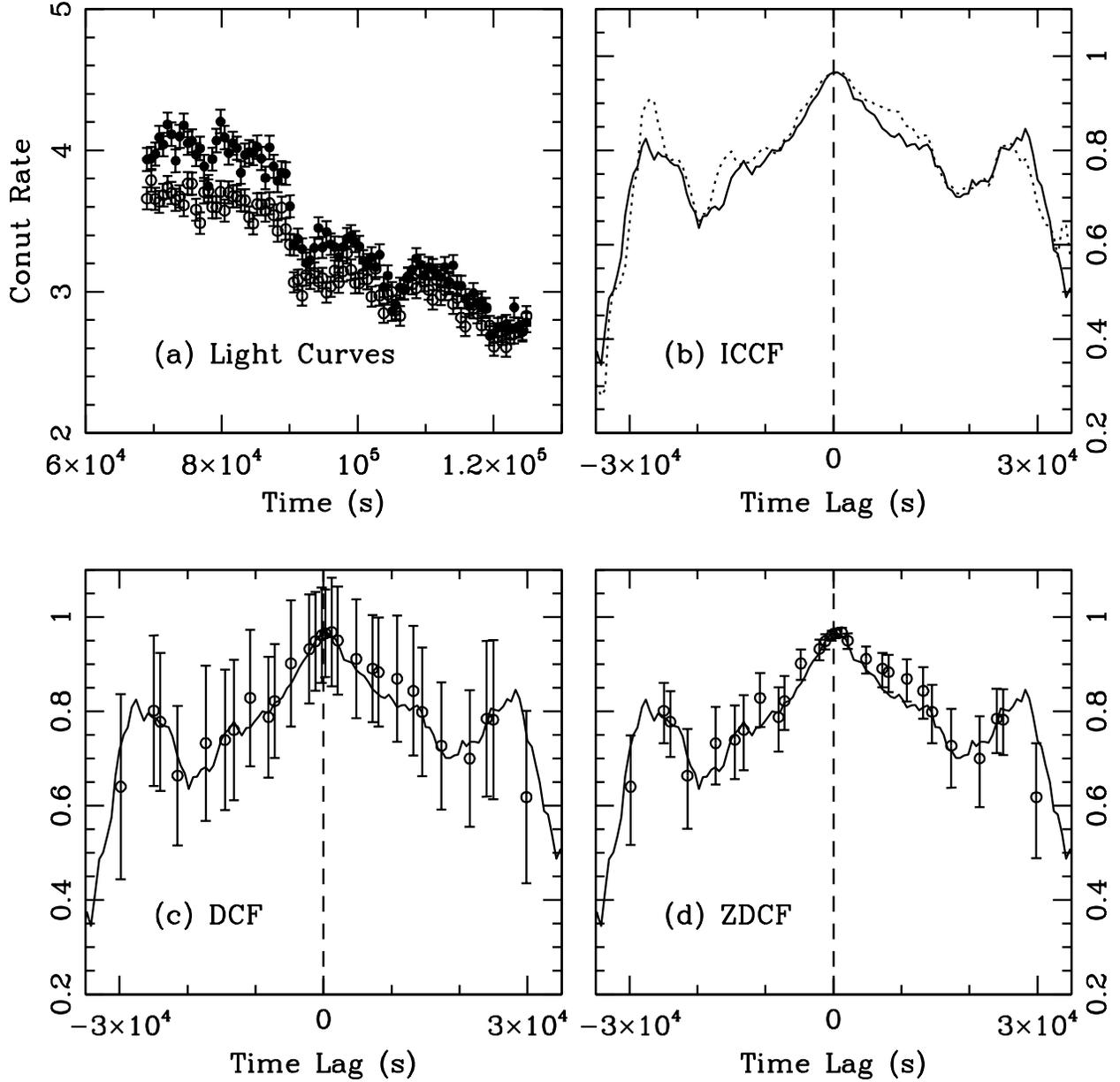}
\caption {\footnotesize (a) A specific case: light curves obtained
with \xmm\ for PKS~2155--304 (see text for details). The solid circles
refer to the 0.7--1~keV energy band, and the open ones to the 1--2~keV
band.  (b) the corresponding ICCF of (a). The solid line is calculated from
real data (i.e., evenly sampled light curves), and the dashed line from
periodically gapped light curves re-sampled from (a) after applying 
the \sax\ sampling windows to (a); (c) the corresponding DCF of (a). The
solid line is calculated from real data, and the open circles with error
bars from the same periodically gapped light curves as used in (b); (d)
same as (c), but for ZDCF.  The dashed lines in (b), (c) and (d) 
indicate zero lag. }
\label{fig:xmmccf}
\end{figure}

In order to perform such a test, we take the first part of the \xmm\
observation of PKS~2155--304 (Maraschi et al. 2002). The details
of the data reduction and CCF analysis will be presented in Maraschi 
et al. (in preparation). For our purposes we just extracted the light curves
in two energy bands, i.e., 0.7-1 and 1-2~keV. The light curves
(without interruptions) are shown in Figure~\ref{fig:xmmccf}a. First, we
performed the CCF analysis on them: the results are shown in
Figure~\ref{fig:xmmccf}b-d (solid line) for the ICCF, DCF and ZDCF, 
respectively. We then re-sampled the light curves with the typical \sax\
sampling windows, obtaining two light curves with periodic 
interruptions resembling real \sax\ observations. We performed the
same CCF analysis on the fake light curves with periodic gaps just as we
did on light curves without gaps. The results are also
shown in Figure~\ref{fig:xmmccf}b-d (dotted line or open circles with
error bars). In all cases the two light curves are highly correlated
near zero lag, and the CCFs calculated from the real and fake data
overlap near their peaks, suggesting that the periodic interruptions do
not change the CCF character near the peak. We also measured the lags with 
the three techniques. The results are tabulated in
Table~\ref{tab:xmmccf}.  Due to the complexities of the CCFs, the
measured lags depend on the CCF methods and the techniques used to
quantify the CCFs. However, the results show in general that the real
lag may be close to zero, with 0.7-1~keV photons very marginally leading 
the 1-2~keV photons.

\begin{figure}
\epsscale{1.0}
\plotone{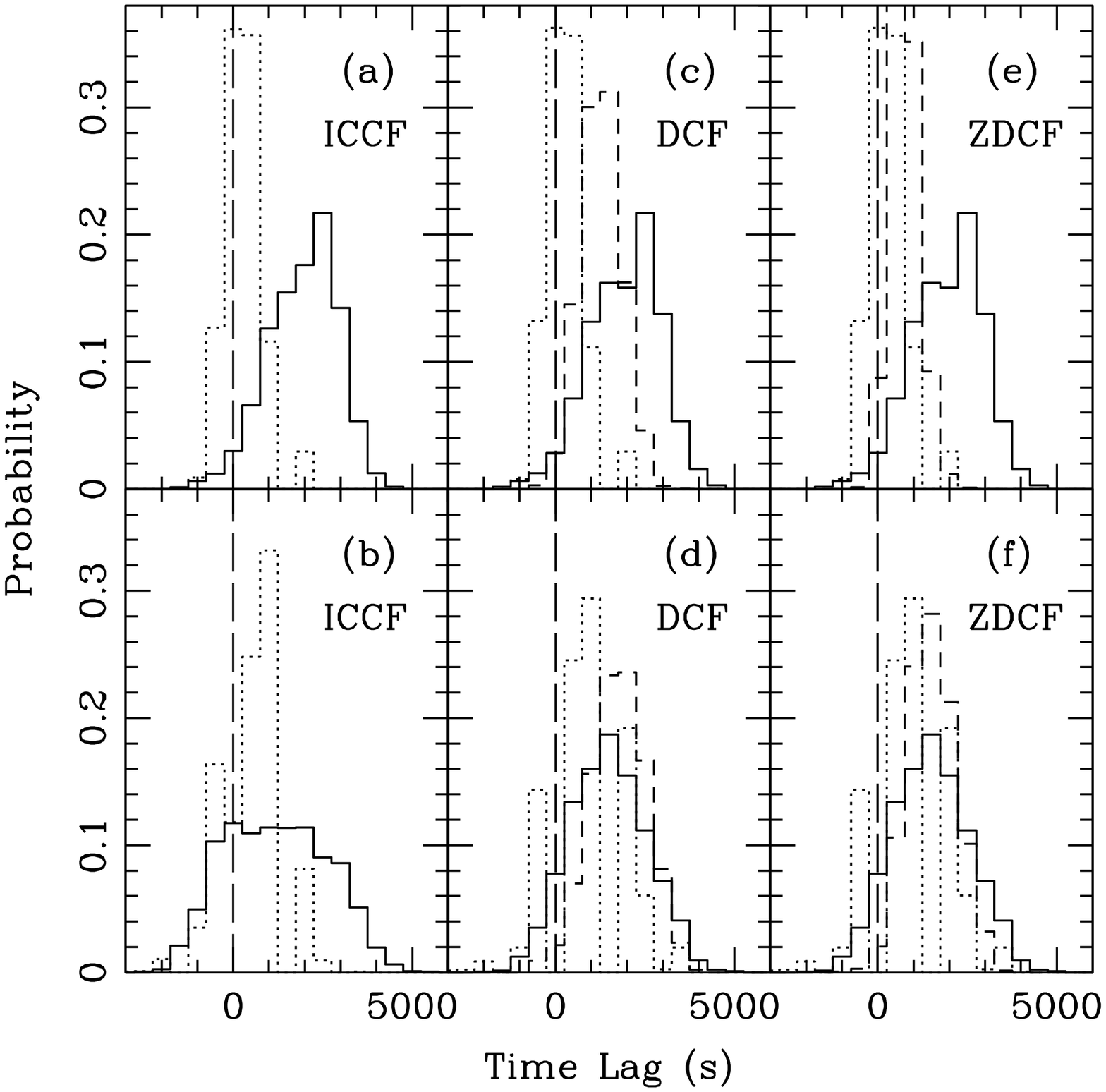}
\caption {\footnotesize CCPDs obtained by Gaussian randomly
redistributing the \xmm\ light curves (Figure~\ref{fig:xmmccf}a) on the
basis of the quoted errors. The upper panels are obtained from the
real \xmm\ light curves, and the lower panels from the
periodically gapped light curves re-sampled from 
the real \xmm\ light curves using the \sax\ sampling windows.  The
dotted line refers to $\tau_{\rm peak}$, the solid line to $\tau_{\rm
cent}$, and the short dashed line to $\tau_{\rm fit}$.  The vertical long
dashed line indicates zero lag.  }
\label{fig:xmmccpd}
\end{figure}

Finally, we performed FR simulations (i.e., Gaussian randomly
redistributing the light curves on the basis of the quoted errors). The
resulting CCPDs are shown in Figure~\ref{fig:xmmccpd}. We used the
same statistical method as we did in \S~\ref{sec:simu:results} to characterize
the CCPDs. The statistical results are tabulated in
Table~\ref{tab:xmmccpd}. Within the $68\%$ confidence errors,
$\tau_{\rm peak}$ gives lags consistent with zero, while $\tau_{\rm cent}$
and $\tau_{\rm fit}$ suggest positive lags of $\sim$1000--2000~s, but at a
low confidence ($\sim$ 2$\sigma$). Note that the low confidence of the lag 
detections may be caused by the complexities of the variability behavior and 
the dependence of variations on the energy band considered. 
This also explains the fact that different methods and techniques may
give rise to inconsistent results. In any case the comparison of 
the results obtained -- using the same method and
technique -- from the light curves with periodic gaps and without gaps does
not favor the suggestion that the periodic interruptions may produce 
spurious lags.     


\section{Discussion and Conclusions}\label{sec:disc}

We performed two sets of simulations to investigate the effects of
space-based observations with short orbital period ($\sim 1.6$ hr) on the
reliability and significance of CCF lag determinations, specifically for
the energy-dependent variations of synchrotron X-ray emission of TeV
blazars. The first set of simulations (\S\ref{sec:simu}) make use of fake
light curves generated with the Fourier transformation method. We
investigated two main issues: (1) the effects of periodic data gaps in the
light curves; and (2) the effects of different lag to bin-size
ratios of light curves. The simulations showed that evenly sampled light 
curves indeed yield more reliable CCF lags, with smaller variance, than the 
periodically gapped light curves do. However, the CCPD analysis clearly 
showed that the light curves with periodic gaps still preserve the nature 
of the true lags (regardless of the values) even though they introduce 
larger lag variances than the light curves without gaps. Moreover, larger 
ratio of lag to light curve bin-size can improve the significance of CCF lag 
determination. The second set of 
simulations is based on a real \xmm\ observation without interruptions. We
re-sampled it with the typical \sax\ sampling windows in order 
to study whether or not periodic interruptions may give rise to strong
biases in CCF lag determinations. The complex nature of the
variability in TeV blazars results in obvious difficulties when quantifying
lags, and it is likely to be the source of discrepancies between the
results quantified with different techniques. However, the comparison
of the results derived with the same CCF method and the same quantifying
technique, shows that the light curves with periodic interruptions do
not produce spurious lags on timescales of hours if their intrinsic 
lag is indeed small. Therefore, our investigations argue against the
proposal by Edelson et al. (2001) that the lags of about an hour discovered
by \asca\ and \sax\ are an artifact of periodic gaps introduced by low
Earth orbit satellites. We thus conclude that the lags discovered by
\asca\ and \sax\ are most likely due to intrinsic variability
properties of the sources, and not artificially produced from an
intrinsic zero lag by periodic gaps. However, due to the 
complexities of variability that produces complicated CCF --
irregularities and complexities of CCF could well be caused by different
properties of variability on different timescales (Zhang 2002) -- it is not
easy to quantify real lags with the CCF method and different CCF
methods and interpreting techniques may work for different cases. Our
simulations also confirm that uninterrupted light curves with large lag to
bin-size ratios can improve accuracy of lag determinations, in particular
for small lags that require high sampling rates of light curves.

Welsh (1999) showed that the reality of CCF lag determinations also depend
(1) on the light curve auto-correlation function (ACF) sharpness (the
sharper the ACFs, the narrower the CCF peak and the smaller the lag bias)
and (2) on the ratio of the intrinsic lag to the duration of the light
curves.  The first dependence can be easily explained by recalling that a
CCF is a convolution of two ACFs. The ACF sharpness is determined by
the PSD steepness (the steeper the PSD, the broader the ACF). The X-ray
PSDs of the three TeV blazars are steep (with slopes of $\sim$ 2--3; Kataoka et
al. 2001; Zhang et al. 1999; 2002), the ACF and the CCF peaks are therefore
broad. De-trending light curves might remove such bias, but it also 
removes the low-(Fourier) frequency variability of the sources. This can 
introduce serious errors into time series and needs to be done very 
carefully.  The second dependence becomes important only for the lengths of 
light curves shorter than $\sim 4$ times the lag, which is never the case for 
the TeV blazars observed in the X-rays (typical ratios of the lengths of light 
curves to lags are 10--100).

Finally, we stress that the real light curves of the bright TeV
blazars are very complex (e.g., the relationship between light curves 
at different energies may not be represented by just one ``fixed''
lag).  Such complexities definitely result in irregular CCF, e.g., the CCF
peaks at zero lag but shows asymmetry, which makes the CCF
methods less straightforward to lag determinations. The
cross-spectral technique, a more complex tool used to determine the Fourier
frequency-dependent lags, may have the advantage of avoiding
such ambiguities at least over long timescales (see Zhang 2002 for
details). However, because this method relies on the Fourier
transformation, it is not applicable to unevenly spaced data.  On the
contrary, in time domain one can use the DCF to substitute the 
classical CCF when dealing with irregular data.

\acknowledgments

We thank the anonymous referee for
constructive comments. This research has made use of the standard on-line
archive provided by the \sax\ Science Data Center (SDC). This work is
partly based on an observation with \xmm , an ESA science mission. The
Italian MIUR is acknowledged for financial support. The project is 
sponsored by the Scientific Research Foundation for the Returned Overseas
Chinese Scholars, State Education Ministry.

\clearpage
\begin{deluxetable}{lccccccc}
\tablecolumns{8}
\tabletypesize{\footnotesize}
\tablewidth{0pt}
\tablecaption{CCF analysis results of one pair of fake light curves(s)\tablenotemark{a}}
\tablehead{
\colhead{} &\multicolumn{3}{c}{Evenly sampled data} 
	 & &\multicolumn{3}{c}{Periodically gapped data}\\
\cline{2-4} \cline{6-8}
\colhead{} 
	&\colhead{$\tau_{\rm peak}$} &\colhead{$\tau_{\rm cent}$}
	&\colhead{$\tau_{\rm fit}$}  &
	&\colhead{$\tau_{\rm peak}$} &\colhead{$\tau_{\rm cent}$}
        &\colhead{$\tau_{\rm fit}$}
}
\startdata
ICCF &$2900$   &$3092$ &... &
     &$2600$   &$2549$ &...\\
DCF  &$3072$   &$3067$  &$3020\pm115$ &
     &$2018$   &$2170$  &$3483\pm359$\\
ZDCF &$3072$   &$3067$  &$3030\pm86$ &
     &$2018$   &$2170$  &$2993\pm287$\\
\enddata
\tablenotetext{a}{The true lag is assumed to be 3000~s.}
\label{tab:simccf}
\end{deluxetable}

\begin{deluxetable}{cccccccc}
\tablecolumns{8}
\tabletypesize{\footnotesize}
\tablewidth{0pt}
\tablecaption{CCPD analysis results of Monte Carlo simulation data (ks)\tablenotemark{a}}
\tablehead{
\colhead{} &\multicolumn{3}{c}{Evenly sampled data} 
	 & &\multicolumn{3}{c}{Periodically gapped data}\\
\cline{2-4} \cline{6-8}
\colhead{True Lag} 
	&\colhead{$\tau_{\rm peak}$} &\colhead{$\tau_{\rm cent}$}
	&\colhead{$\tau_{\rm fit}$}  &
	&\colhead{$\tau_{\rm peak}$} &\colhead{$\tau_{\rm cent}$}
        &\colhead{$\tau_{\rm fit}$}
}
\startdata
     &\multicolumn{7}{c}{ICCF} \\
%
0.00 &$0.10^{+0.10}_{-0.30}$ &$0.00^{+0.11}_{-0.10}$ &... &
     &$0.10^{+0.10}_{-0.30}$ &$0.00^{+0.24}_{-0.24}$ &...\\
0.30 &$0.30^{+0.40}_{-0.10}$ &$0.31^{+0.22}_{-0.20}$ &... &
     &$0.30^{+0.40}_{-0.10}$ &$0.30^{+0.24}_{-0.25}$ &...\\
1.40 &$1.30^{+0.10}_{-0.10}$ &$1.40^{+0.15}_{-0.14}$ &... &
     &$1.30^{+0.40}_{-0.10}$ &$1.32^{+0.42}_{-0.35}$ &...\\
3.00 &$2.90^{+0.40}_{-0.10}$ &$3.00^{+0.19}_{-0.19}$ &... &
     &$3.00^{+0.30}_{-0.30}$ &$2.99^{+0.53}_{-0.47}$ &...\\
5.40 &$5.40^{+0.50}_{-0.10}$ &$5.40^{+0.24}_{-0.24}$ &... &
     &$5.40^{+0.40}_{-0.10}$ &$5.44^{+0.30}_{-0.32}$ &...\\
7.10 &$7.00^{+0.40}_{-0.10}$ &$7.10^{+0.27}_{-0.24}$ &... &
     &$7.10^{+0.40}_{-0.10}$ &$7.03^{+0.48}_{-0.38}$ &...\\
     &\multicolumn{7}{c}{DCF} \\

0.00 &$0.00^{+0.00}_{-0.00}$ &$0.00^{+0.23}_{-0.23}$ &$0.10^{+0.18}_{-0.16}$ &
     &$0.00^{+0.00}_{-0.00}$ &$0.00^{+0.29}_{-0.07}$ &$0.00^{+0.46}_{-0.44}$\\
0.30 &$0.51^{+0.00}_{-0.00}$ &$0.26^{+0.23}_{-0.02}$ &$0.30^{+0.18}_{-0.17}$ &
     &$0.51^{+0.00}_{-0.51}$ &$0.05^{+0.78}_{-0.05}$ &$0.29^{+0.47}_{-0.45}$\\
1.40 &$1.54^{+0.00}_{-0.00}$ &$1.31^{+0.23}_{-0.02}$ &$1.38^{+0.28}_{-0.23}$ &
     &$1.02^{+0.00}_{-0.00}$ &$1.14^{+0.65}_{-0.45}$ &$1.36^{+0.67}_{-0.62}$\\
3.00 &$3.07^{+0.00}_{-0.00}$ &$3.06^{+0.24}_{-0.24}$ &$2.96^{+0.44}_{-0.38}$&
     &$2.02^{+0.00}_{-0.00}$ &$2.47^{+0.74}_{-0.41}$ &$2.98^{+0.85}_{-0.79}$\\
5.40 &$5.63^{+0.00}_{-0.51}$ &$5.38^{+0.25}_{-0.24}$ &$5.34^{+0.69}_{-0.64}$ &
     &$6.14^{+0.00}_{-1.54}$ &$4.93^{+0.95}_{-0.72}$ &$5.35^{+0.78}_{-0.62}$\\
7.10 &$7.17^{+0.00}_{-0.00}$ &$7.16^{+0.25}_{-0.24}$ &$7.02^{+0.90}_{-0.81}$ &
     &$6.66^{+0.76}_{-0.00}$ &$6.91^{+0.72}_{-0.93}$ &$7.03^{+0.94}_{-0.82}$\\
%
     &\multicolumn{7}{c}{ZDCF} \\
%
0.00 &$0.00^{+0.00}_{-0.00}$ &$0.00^{+0.23}_{-0.23}$ &$0.00^{+0.10}_{-0.10}$ &
     &$0.00^{+0.00}_{-0.00}$ &$0.00^{+0.29}_{-0.07}$ &$0.03^{+0.25}_{-0.23}$\\
0.30 &$0.51^{+0.00}_{-0.00}$ &$0.26^{+0.23}_{-0.02}$ &$0.31^{+0.10}_{-0.10}$ &
     &$0.51^{+0.00}_{-0.51}$ &$0.05^{+0.78}_{-0.05}$ &$0.34^{+0.25}_{-0.25}$\\
1.40 &$1.54^{+0.00}_{-0.00}$ &$1.31^{+0.23}_{-0.02}$ &$1.40^{+0.13}_{-0.12}$ &
     &$1.02^{+0.00}_{-0.00}$ &$1.14^{+0.65}_{-0.45}$ &$1.43^{+0.37}_{-0.38}$\\
3.00 &$3.07^{+0.00}_{-0.00}$ &$3.06^{+0.24}_{-0.24}$ &$3.00^{+0.18}_{-0.19}$&
     &$2.02^{+0.00}_{-0.00}$ &$2.47^{+0.74}_{-0.41}$ &$2.97^{+0.61}_{-0.54}$\\
5.40 &$5.63^{+0.00}_{-0.51}$ &$5.38^{+0.25}_{-0.24}$ &$5.38^{+0.26}_{-0.32}$ &
     &$6.14^{+0.00}_{-1.54}$ &$4.93^{+0.95}_{-0.72}$ &$5.49^{+0.48}_{-0.39}$\\
7.10 &$7.17^{+0.00}_{-0.00}$ &$7.16^{+0.25}_{-0.24}$ &$7.08^{+0.30}_{-0.41}$ &
     &$6.66^{+0.76}_{-0.00}$ &$6.91^{+0.72}_{-0.93}$ &$7.13^{+0.49}_{-0.47}$\\
\enddata
\tablenotetext{a}{The quoted values are the medians of the CCPDs, and the
errors are $68\%$ confidence range with respect to the 
medians.
\\
}
\label{tab:simccpd}
\end{deluxetable}

\begin{deluxetable}{lccccccc}
\tablecolumns{8}
\tabletypesize{\footnotesize}
\tablewidth{0pt}
\tablecaption{CCF analysis results of the \xmm\ data (s)}
\tablehead{
\colhead{} &\multicolumn{3}{c}{Real data} 
	 & &\multicolumn{3}{c}{Fake data}\\
\cline{2-4} \cline{6-8}
\colhead{} 
	&\colhead{$\tau_{\rm peak}$} &\colhead{$\tau_{\rm cent}$}
	&\colhead{$\tau_{\rm fit}$}  &
	&\colhead{$\tau_{\rm peak}$} &\colhead{$\tau_{\rm cent}$}
        &\colhead{$\tau_{\rm fit}$}
}
\startdata
ICCF &$300$   &$1904$ &... &
     &$300$   &$-212$ &...\\
DCF  &$0$     &$100$  &$695\pm865$ &
     &$1200$  &$707$  &$1538\pm2547$\\
ZDCF &$0$     &$100$  &$433\pm179$ &
     &$1200$  &$707$  &$879\pm592$\\
\enddata
\label{tab:xmmccf}
\end{deluxetable}

\begin{deluxetable}{lccccccc}
\tablecolumns{8}
\tabletypesize{\footnotesize}
\tablewidth{0pt}
\tablecaption{CCPD analysis results of the \xmm\ data (s)\tablenotemark{a}}
\tablehead{
\colhead{} &\multicolumn{3}{c}{Real data} 
	 & &\multicolumn{3}{c}{Fake data}\\
\cline{2-4} \cline{6-8}
\colhead{} 
	&\colhead{$\tau_{\rm peak}$} &\colhead{$\tau_{\rm cent}$}
	&\colhead{$\tau_{\rm fit}$}  &
	&\colhead{$\tau_{\rm peak}$} &\colhead{$\tau_{\rm cent}$}
        &\colhead{$\tau_{\rm fit}$}
}
\startdata
ICCF &$600^{+600}_{-600}$  &$2021^{+867}_{-1092}$ &... &
     &$600^{+600}_{-1200}$ &$1201^{+1617}_{-1541}$ &...\\
DCF  &$0^{+0.600}_{-600}$ &$2011^{+881}_{-1104}$ &$1279^{+566}_{-557}$&
     &$1200^{+600}_{-1800}$ &$1477^{+1101}_{-1110}$ &$1767^{+777}_{-787}$\\
ZDCF &$0^{+0.600}_{-600}$ &$2011^{+881}_{-1104}$ &$712^{+418}_{-358}$&
     &$1200^{+600}_{-1800}$ &$1477^{+1101}_{-1110}$ &$1457^{+716}_{-623}$\\
\enddata
\tablenotetext{a}{The quoted values are the medians of the CCPDs, and the
errors are $68\%$ confidence range with respect to the 
medians.
\\
}
\label{tab:xmmccpd}
\end{deluxetable}

\clearpage

\clearpage

\clearpage

\clearpage

\clearpage

\clearpage

\end{document}